\documentclass[journal=nanolett,manuscript=article]{achemso}

\pdfoutput=1
\usepackage{setspace}
\usepackage{graphicx}
\usepackage{rotating}
\usepackage{amsmath, amssymb, amsthm}
\usepackage{color}
\usepackage[colorinlistoftodos]{todonotes}

\author{Lei Liu}
\affiliation{Korea Institute for Advanced Study, Seoul 02455, Korea}
\author{Philip A. Pincus}
\affiliation{Physics and Material Department, University of California, Santa Barbara, U. S. A.}
\author{Changbong Hyeon}
\email{hyeoncb@kias.re.kr}
\affiliation{Korea Institute for Advanced Study, Seoul 02455, Korea}

\title{Compressing $\Theta$-chain in slit geometry}

\begin{document}

\begin{tocentry}
\centering
\includegraphics[width=0.55\textwidth]{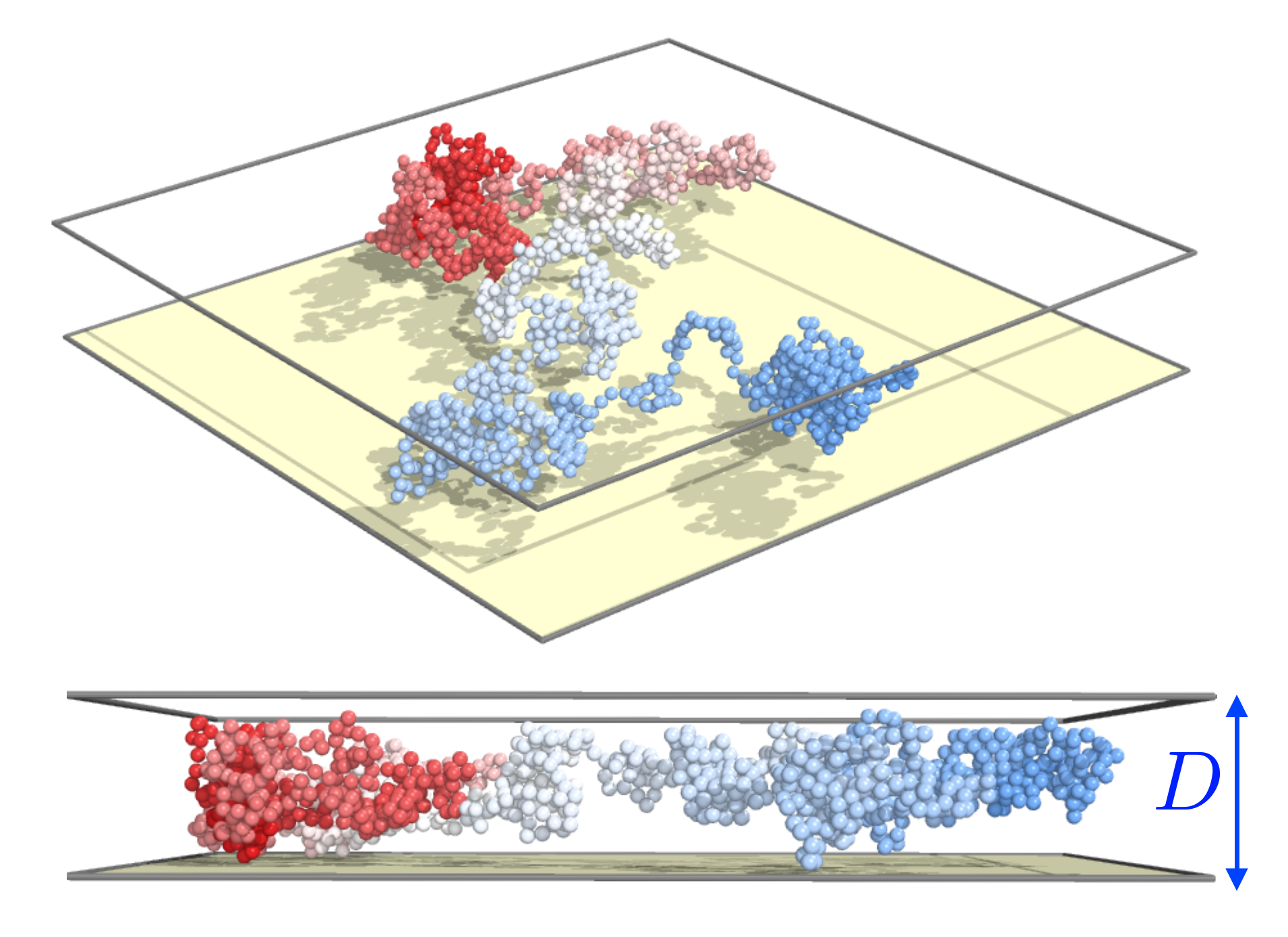}
\end{tocentry}

\begin{abstract}
When compressed in a slit of width $D$, 
a $\Theta$-chain that displays the scaling of size $R_0$ (diameter) with respect to the number of monomers $N$, $R_0\sim aN^{1/2}$, expands in the lateral direction as 
$R_{\parallel}\sim a N^{\nu}(a/D)^{2\nu-1}$. 
Provided that the $\Theta$ condition is strictly maintained throughout the compression, 
the well-known scaling exponent of $\Theta$-chain in 2 dimensions, $\nu=4/7$, is anticipated in a perfect confinement.  
However, numerics shows that upon increasing compression from $R_0/D<1$ to $R_0/D\gg 1$, $\nu$ gradually deviates from $\nu=1/2$ and plateaus at $\nu=3/4$, 
the exponent associated with the self-avoiding walk in two dimensions. 
Using both theoretical considerations and numerics, 
we argue that it is highly nontrivial to maintain the $\Theta$ condition under confinement because of two major effects. 
First, as the dimension is reduced from 3 to 2 dimensions, 
the contributions of higher order virial terms, which can be ignored in 3 dimensions at large $N$, become significant. 
Second and more importantly, the geometrical confinement, which is regarded as an applied external field, alters the second virial coefficient ($B_2$) changes  
from $B_2=0$ ($\Theta$ condition) in free space to $B_2>0$ (good-solvent condition) in confinement. 
Our study provides practical insight into how confinement affects the conformation of a single polymer chain. 
\end{abstract}
\maketitle 

\section{Introduction}
The structural and dynamical properties of flexible polymer chains confined under a geometrical constraint are relevant to a number of problems in biology and material sciences.  
Whereas there have been a plethora of experimental and theoretical studies on the effects of various geometrical constraints on self-avoiding polymer \cite{Daoud75Macromolecules,Daoud77JP,tegenfeldt2004PNAS,hsu2004JCP,Morrison05JCP,dimitrov2008JCP,Kang2015PRL,Ha2015SoftMatter} and collapse/folding transitions of biopolymers \cite{zhou2001Biochem,obrien2008NL}, 
relatively less attention has been paid to a single $\Theta$-chain upon confinement \cite{hsu2005JSM,cifra2000concentration,cifra2003macromolecules}. 

The $\Theta$ condition is a point at which the pair-wise repulsion and attraction between monomers are balanced so that the second virial coefficient vanishes. 
De Gennes \cite{deGennes75JPL} has identified that the $\Theta$ point of polymer system corresponds to the tricritical point of thermodynamic system, which is best exemplified by He$^3$-He$^4$ mixtures where the continuous, second-order phase transition between super-fluid and normal phases along $\lambda$-line ceases and undergoes phase separation \cite{Griffiths1970PRL,blume1971PRA}. 
Such conditions are well known in biological as well as polymer contexts. 
In polymer solution under semi-dilute to concentrated condition, the radius of gyration of a chain molecule scales with $N^{1/2}$.
In studies on chromatin conformation inside a cell nucleus, there is evidence that the ``subchains" in certain cell types or organisms form an \emph{equilibrium globule} that is suspected to lie close to a $\Theta$ condition, 
yielding $R(s)\sim s^{1/2}$ where $s$ is the length of the subchain. \cite{Halverson14RPP,Kang2015PRL,liu2016BJ} 
Raphael and Pincus \cite{Raphael92JP} studied the size of a single $\Theta$-chain with varying length ($N$) when the chain is confined in a cylindrical pore geometry. 
Our problem of interest in this study is the two-dimensional version of Ref.\cite{Raphael92JP}. 
We study the effect of compression, in a slit geometry, of the size of a polymer chain that is originally prepared under $\Theta$ conditions without geometrical constraint. 

Experimentally the scaling exponent for 2D polymer in $\Theta$-solvent condition ($R_\parallel\sim N^{\nu_2}$) was measured for the first time by Vilanove and Rondelez who used the pressure isotherms of thin polymer-film in the high-concentration regime ($c^*\ll c\ll 1$) \cite{vilanove1980PRL},  
where $c^*\sim N/R_{\parallel}^2D$ is the critical overlap concentration.  
For concentrations above the semi-dilute condition ($c>c^*$), the osmotic pressure in 2 dimensions is expected to show a concentration dependence of $\frac{\Pi}{T}\sim \frac{c}{N}f\left(\frac{c}{c^*}\right)$.  
Above the semi-dilute concentration, the osmotic pressure is determined by the local density of monomers only, not depending on the length ($N$) of individual chains. 
This condition allows us to determine 
the scaling relation of $\frac{\Pi}{T}\sim c^{\alpha}$ with $\alpha=\frac{2\nu_2}{2\nu_2-1}$.  
By experimenting surface-pressure isotherms in Langmuir monolayers of polymethylmethacrylate (PMMA), Vilanove and Rondelez obtained $\nu_2=0.56\pm 0.01$ \cite{vilanove1980PRL}.  
More recently, Witte \emph{et al.}\cite{witte2010macromolecules} found using the same technique that $\nu_2=0.57\pm 0.02$ for poly($n$-butyl acrylate).

Historically, there were a number of theoretical studies to determine the fractal dimension of the ``hull" (external perimeter) of a cluster ($D_H$) at the percolation threshold ($p_c=1/2$) in 2 dimensions \cite{Bunde85JPA,Weinrib85PRB,Grassberger86JPA}. 
Coniglio \emph{et al}  \cite{coniglio1987PRB} posited that the hull exponent ($\nu=D_H^{-1}=4/7\approx 0.5714$) of a percolating cluster is equivalent to the size exponent of a polymer chain in 2 dimensions at the coil-globule transition ($\Theta'$) point.  
Duplantier and Saleur obtained the exact tricritical point ($\Theta$) and associated exponents ($\nu=4/7$, $\gamma=8/7$, and $\phi=3/7$) in two dimensions using a model of self-avoiding walk on a honeycomb lattice with percolating vacancies \cite{Duplantier87PRL}. 
Later, they used a method of conformal invariance and showed that the hypothesized $\Theta'$ point coincides with the tricritical $\Theta$ point of a polymer in two dimensions \cite{duplantier1989PRL}.

Although it appears that the problem of $\Theta$-chain in slit confinement is mostly resolved, 
there still remain practical issues. 
First, the  $\Theta$ condition is a singular point, which is not easy to attain for a single chain. 
More seriously, the magnitude of virial coefficients not only depend on the solvent quality or temperature, but also on an external field \cite{Glandt80JCIS,Yang2015MolPhys,Krekelberg2019JCP}. 
A geometrical constraint, which can be regarded as a type of external field, indeed alters 
the virial coefficients. 
Second, the $\Theta$ condition itself in two dimensions 
gives another subtlety to the problem. 
The contributions of higher order virial terms is of the same order of magnitude as the lower order terms. 
In other words, the perturbative expansion, usually employed in a Flory-type argument, is no longer applicable for describing $\Theta$-chain configurations in $d\leq 2$ dimensions. 

In this study, we first review the conventional scaling and blob arguments for polymer chains under confinement and apply them to $\Theta$-chains. 
Second, we revisit the Flory argument with which to determine polymer size and discuss its limitations, which is directly associated with the problem studied here. 
Third, we present our simulation results for a three dimensional $\Theta$-chain that is compressed into a slit.   The
$\Theta$ condition being strictly maintained, it is expected that the size scaling exponent of a polymer chain undergoes crossover from $\nu=1/2$ to $\nu=4/7$ upon perfect confinement in 2 dimensions. 
Instead, the value of exponent reaches $\nu=3/4$ when the chain is sufficiently compressed.  
We then discuss the origin of this behavior.

\section{Theoretical considerations}

{\bf Scaling argument.} A single chain, immersed in a bulk $\Theta$-solvent, approximately behaves like an ideal chain, obeying the the scaling relation of size $R_0$ with the number of monomers:  
\begin{align}
R_0\sim aN^{1/2}. 
\end{align}
where $R_0\sim 2 R_g$ with the radius of gyration $R_g$. 
The chain is now confined between perfectly repelling walls of gap $D$. 
Then, a scaling form of $R_\parallel$ is given as 
\begin{align}
R_\parallel\sim R_0f(R_0/D). 
\end{align}
If $D\gg R_0$, there is no change in the size of the chain: the chain remains ideal, thus $f(x)\sim 1$, and $R_\parallel\sim R_0$. 
But, if $D\ll R_0$, the chain is squeezed into a pancake of lateral dimension $R_{\parallel}$. 
Then, $R_{\parallel}$ is expected to scale as 
$R_{\parallel}\sim R_0f(R_0/D)\sim R_0(R_0/D)^m$. 
Using the relation of $R_0\sim a N^{1/2}$ and $\nu=\frac{1}{2}(1+m)$, we obtain the scaling relation between $R_\parallel$ and $N$. 
\begin{align}
R_{\parallel}\sim 
aN^{\nu}\left(\frac{a}{D}\right)^{2\nu-1}. 
\label{eqn:scaling}
\end{align}
In the above scaling relation, the precise value of the exponent $\nu$ is not specified here; however, the relation of $R_\parallel\sim N^{\nu}$ constrains the scaling relation between $R_\parallel$ and $D$ into  
$R_\parallel\sim D^{1-2\nu}$.
\\

{\bf Blob argument.} Alternatively, the blob argument may be used to derive the same scaling relation (Eq.\ref{eqn:scaling}). 
Upon confinement with $R_0>D$, the chain is partitioned into $N/g$ blobs where $g$ is the number of monomers comprising a single blob, whose size should be comparable to $D$. 
As long as the size of chain is large enough, the monomers inside a blob should still be in the same solvent condition ($\Theta$ condition). 
Thus, it can be argued that $g$ monomers comprising each blob still obey the ideal chain statistics.
\begin{align}
D\sim ag^{1/2}.
\label{eqn:blob1} 
\end{align} 
Given that there are $N/g$ blobs, the lateral size of confined polymer is 
\begin{align}
R_\parallel\sim D\left(\frac{N}{g}\right)^{\nu}. 
\label{eqn:blob2}
\end{align}
Eliminating $g$ between Eqs.~\ref{eqn:blob1} and ~\ref{eqn:blob2}, we obtain the scaling relation  (Eq.\ref{eqn:scaling}). 
Notice that for given parameters, $N$, $a$, and $D$, the confined chain is partitioned into $N/g=N(a/D)^2$ blobs. 

\section*{Effective energy hamiltonian and Flory argument for a polymer chain}
 The effective energy hamiltonian in $d$-dimension (Edwards hamiltonian \cite{edwards1979JCSFaraday}) that describes a polymer chain is given  by 
\begin{align}
\beta H_{\text{eff}}&=\frac{1}{2}\int_0^L\left(\frac{\partial {\bf r}}{\partial s}\right)^2ds+\frac{B_2}{2!}\int_0^Lds\int_0^Lds'\delta^d[{\bf r}(s)-{\bf r}(s')]\nonumber\\
&+\frac{B_3}{3!}\int_0^Lds\int_0^Lds'\int_0^Lds''\delta^d[{\bf r}(s)-{\bf r}(s')]\delta^d[{\bf r}(s')-{\bf r}(s'')]+\cdots 
\end{align}
where the first term describes the elastic deformation arising from chain connectivity. 
The second and third terms for two body and three body interaction potentials; and the coefficients $B_2$ and $B_3$ are the virial coefficients with dimension of $[B_2]\sim a^3$ and $[B_3]\sim a^6$. 
The partition function for the chain is written as
\begin{align}
Z&=e^{-\beta F}= \int \mathcal{D}[{\bf r}(s)]e^{-\beta H_{\text{eff}}}\sim \langle e^{-\beta H_{\text{eff}}}\rangle\geq e^{-\beta \langle H_{\text{eff}}\rangle},
\end{align} 
where $\langle\ldots\rangle=\frac{\int\mathcal{D}[{\bf r}(s)](\ldots)}{\int\mathcal{D}[{\bf r}(s)]}$ is the average over all realizations of the  polymer configurations, and the last relation employs Jensen's inequality for convex functions. 
Hence, the free energy of the polymer chain is obtained by minimizing the effective energy hamiltonian.  
\begin{align}
\beta F&\leq \beta \langle H_{\text{eff}}\rangle
\end{align}
Employing the density (or concentration) fields, $c(R)=\int_0^Lds\delta^d[{\bf r}(s)-R]$,  the pairwise interaction may be rewritten as \cite{desCloizeauxBook}, 
\begin{align}
\int_0^Lds\int_0^Lds'\delta^{d}[{\bf r}(s)-{\bf r}(s')]&=\int d^dR\underbrace{\int_0^Lds\delta^{d}[{\bf r}(s)-R]}_{=c(R)}\underbrace{\int_0^Lds'\delta^{d}[{\bf r}(s')-R]}_{=c(R)}\nonumber\\
&=\int d^dR c^2(R)
\end{align}
Thus, the average of energy hamiltonian is given by 
\begin{align}
\beta F\leq \beta \langle H_{\text{eff}}\rangle&=\frac{1}{2}\Big\langle\int_0^L\left(\frac{\partial {\bf r}}{\partial s}\right)^2ds\Big\rangle+\frac{B_2}{2!}\int d^dR\langle c^2(R)\rangle+\frac{B_3}{3!}\int d^dR\langle c^3(R)\rangle+\cdots \nonumber\\
&\approx \frac{1}{2}\left(\frac{R^2}{Na^2}\right)+\frac{B_2}{2!}
\langle c\rangle^2 R^d+\frac{B_3}{3!}\langle c\rangle^3 R^d+\cdots \nonumber\\
&\approx \frac{1}{2}\frac{R^2}{Na^2}+\frac{B_2}{2!}\frac{N^2}{R^d}+\frac{B_3}{3!}\frac{N^3}{R^{2d}}+\cdots 
\label{eqn:Flory}
\end{align}
where the approximation $\langle c^n\rangle\approx \langle c\rangle^n$ is used. 
This rationalizes the Flory-type free energy for a polymer chain. 

The Flory free energy of a confined chain in a slit is given by 
\begin{align}
\beta F\sim \frac{R_{\parallel}^2}{Na^2}+\frac{B_2}{2!} c^2V+\frac{B_3}{3!}c^3V+\cdots\frac{B_n}{n!}c^nV+\cdots 
\end{align}
with the monomer density $c\sim N/V$  and the volume of the confined polymer given by $V\sim R_\parallel^2D$. 
At good-solvent condition, $B_2\sim \tau a^3>0$ with $\tau=(T-T_{\Theta})/T_{\Theta}$. 
A marginal solvent close to the $\Theta$ condition has $0\lesssim \tau<1$. 
The Flory free energy of a confined chain is 
\begin{align}
\beta F\sim \frac{R_{\parallel}^2}{Na^2}+\frac{B_2}{2!}\frac{N^2}{R^2_\parallel D}+\cdots 
\end{align}
Minimization of this free energy with respect to $R_\parallel$ leads to 
$R^{\text{SAW}}_\parallel\sim \tau aN^{3/4}(a/D)^{1/4}$. 

On the other hand, at the $\Theta$ point, $B_2\sim \tau a^3=0$ because of $\tau=(T-T_{\Theta})/T_{\Theta}=0$ \cite{deGennes75JPL}. 
In this case, it is tempting to consider the Flory free energy of the confined chain as 
\begin{align}
\beta F\sim  \frac{R_{\parallel}^2}{Na^2}+\frac{B_3}{3!}\frac{N^3}{(R^2_\parallel D)^2}+\cdots. 
\end{align}
While minimization of this free energy gives rise to $R^{\Theta_F}_\parallel\sim aN^{2/3}(a/D)^{1/3}$, the exponent $\nu_F=2/3\approx 0.667$ differs significantly from the exact value $\nu=4/7\approx 0.571$ \cite{Duplantier87PRL}. 

The above type of argument using the Flory free energy requires extra care and in fact cannot be applied for studying $\Theta$-chain in $d\leq 2$ because of the following.  
In Eq.\ref{eqn:Flory}, $R\sim N^{1/2}$ near $\Theta$ condition gives rise to the higher order virial terms that scale with $N$ as 
\begin{align}
F_2&\sim B_2N^{2-d/2}\nonumber\\
F_3&\sim B_3N^{3-d}\nonumber\\
F_n&\sim B_nN^{n-(n-1)d/2}
\end{align}

This consideration engenders several key messages: 

(i) In 3 dimensions ($d=3$) and for $N\gg 1$, the second virial term is the most dominant in the entire virial series ($F_2\sim N^{1/2}\gg F_3\sim N^0\gg F_n$). -- more precisely, for $\Theta$-chain ($R^2\sim a^2N$), the third virial term yields a logarithmic contribution to the free energy ($F_3\sim 4\pi B_3\int_{R_{\text{min}}}^{R_{\text{max}}}c^3R^2dR\sim (2\pi B_3/a^6)\log{N}$) --  For this reason, it suffices in 3 dimensions to consider the second virial coefficient to assess the condition of the $\Theta$-solvent for a chain with large $N$ unless the second virial term very nearly vanishes.

(ii) Even in the case $B_2(T)$ precisely vanishes ($B_2(T)=0$, tricritical point), the higher order virial terms still contribute to determining the polymer configuration. 
Specifically, the third virial term gives a logarithmic correction to the free energy. 
It was shown that the gyration radius of $\Theta$-chain in 3 dimensions is 
\begin{align}
\frac{R_0^2}{Na^2}\sim A_0(y)\left(1-\frac{493\pi}{33\times 4}\frac{y}{1+44\pi y\log{N}}\right)
\label{eqn:higher_order}
\end{align}
where $A_0(y)=1+\frac{16}{33}\pi y+\cdots$ with $y=(2\pi)^{-3}B_3$ \cite{Duplantier87JCP}.

(iii) In 2 dimensions ($d=2$), all the virial terms are comparable in magnitude, such that $F_2\sim F_3\sim \cdots$, i.e., $F_n\sim N$ for all $n$'s. 
Therefore, the perturbative expansion is inherently problematic for $d\leq 2$. 

(iv) Raphael and Pincus obtained the scaling of $\Theta$-chain size with $N$ in cylindrical confinement for $R_0/D\ll 1$, namely $R_\parallel\sim aN(a/D)$, by considering the Flory free energy of elastic term and the third virial term. Interestingly, for the 1D confinement problem, the expected scaling relationship of $R_\parallel\sim N$ can be derived by balancing the elastic term and any $n$-th virial term, such that 
\begin{align}
\beta F\sim \frac{R_\parallel^2}{Na^2}+\frac{B_n}{n!}\frac{N^n}{(R_\parallel D^2)^{n-1}}, 
\end{align}
with $[B_n]\sim a^{3(n-1)}$. 
The minimization of $\beta F$ with respect to $R_\parallel$ gives 
\begin{align}
R_\parallel\sim aN\left(\frac{a}{D}\right)^{\frac{2(n-1)}{n+1}}. 
\end{align} 
However, Eq.\ref{eqn:scaling} indicates that the only $n$-th virial term that gives rise to the correct scaling of $R_\parallel\sim D^{-1}$ is when $n=3$.

\section{Results from numerics}
\begin{figure*}[t]
\centering
\includegraphics[width=0.6\textwidth]{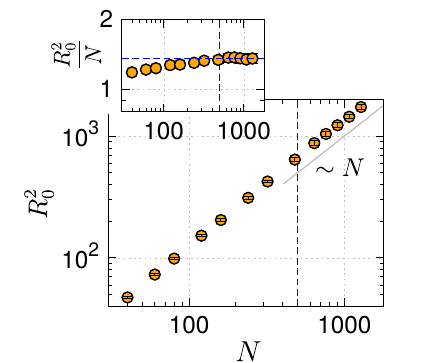}
\caption{Scaling of a $\Theta$ polymer ($B_2\approx 0$ at $\epsilon = 0.335$ $k_BT$) in 3 dimensions. 
It is expected that $R_0^2/N\sim N^0$ under $\Theta$ conditions but this relation is strictly realized only at large $N$. For small $N$, the chain is swollen with $R^2_0/N\sim N^{\alpha}$ ($\alpha>0$).}
\label{fig1}
\end{figure*}

In order to generate a $\Theta$-chain in bulk,  a string of Lennard Jones (LJ) particles was used. 
The second virial coefficient between monomers was made to nearly vanish ($B_2=1.76\times 10^{-3}a^3$) by choosing the depth of the LJ potential as $\epsilon=0.335$ $k_BT$ (see Method for details). 
Note that to observe the correct scaling of a $\Theta$-chain, $N$ should be sufficiently large. 
Consistent with Eq.\ref{eqn:higher_order}, the simulation result for the polymer size ($R_0$, the diameter of polymer calculated as $R_0=2\times R_g$) at small $N$ shows deviation from the ideal chain behavior. 
In our numerics, the ideal regime ($R_0^2/N\sim \text{const.}$) is attained for $N\gtrsim 500$ (Fig.\ref{fig1}). 
Although not explicitly captured by our simulation, because of the finite size $N(\lesssim 10^3)$, another crossover point from ideal polymer to SAW at large $N$ is expected for $d=3$ because $\tau$ is not exactly zero ($\tau=1.76\times 10^{-3}$).  It is expected that from the condition of $\tau N^{2-d/2}>N^{3-d}$ that the SAW statistics  emerges when  
\begin{align}
N>\tau^{-2}. 
\end{align} 
This implies that the chain in a marginal solvent close to $\Theta$ point ($\tau\gtrsim 0$) behaves as an ideal chain if $N$ is smaller than $\tau^{-2}$; but displays SAW statistics when $N>\tau^{-2}$. 
  
\begin{figure*}[t]
\centering
\includegraphics[width=0.9\textwidth]{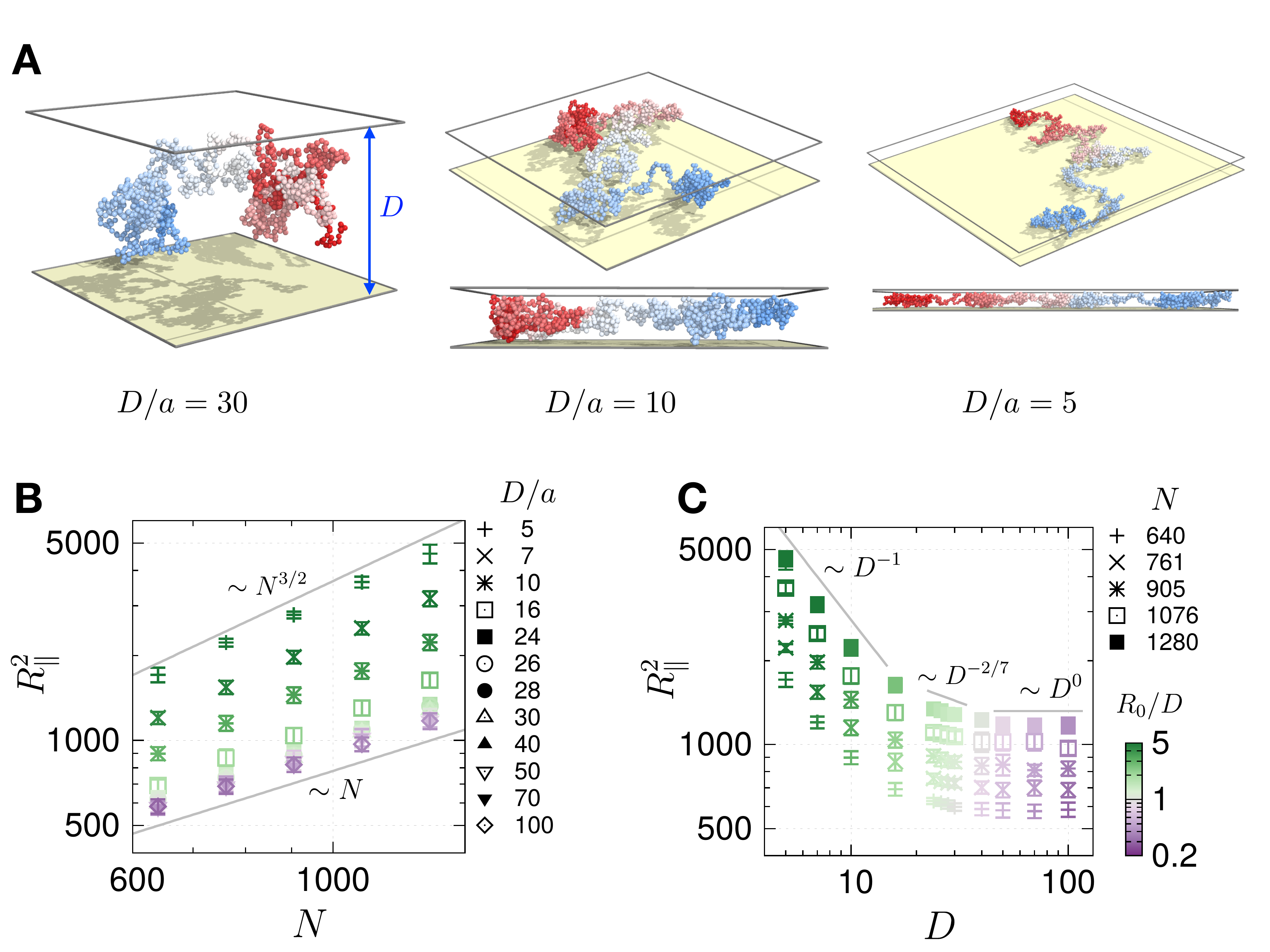}
\caption{
(A) Snapshots of polymer configurations ($N=1280$) under confinement from simulations.  
(B) $\Theta$-chains that  in 3 dimensions obey $R_0\sim N^{1/2}$ change the scaling of their lateral dimension 
with $N$ ($R_\parallel\sim N^{\nu}$) from $\nu=1/2$ to $\nu=3/4$ with decreasing $D$ (increasing compression). 
(C) The scaling of $R_\parallel$ with respect to $D$ ($R_\parallel^2 \sim D^{-2(2\nu-1)}$) is also explained with the variation of $\nu$ from $\nu=1/2$ to $\nu=3/4$ through $\nu=4/7$ with increasing compression (from $R_0/D\gtrsim 1$ to $R_0/D\gg 1$).  
In (B) and (C), the color of the data points represents the extent of confinement quantified by $R_0/D$. 
The purple color means that chain is effectively in free space, whereas the green color means that chain is confined.  
}
\label{fig2}
\end{figure*}

Next,  the scaling of lateral size of polymer ($R_\parallel$) as the polymer of size $N>500$ is compressed from $R_0/D\gtrsim 1$ to $R_0/D\gg 1$ was examined (Fig.\ref{fig2}). 
As shown in Fig.\ref{fig2}B, the scaling relation of $R_{\parallel}^2\sim N$ is obeyed when the chain is effectively free from confinement ($R_0<D\simeq 100$).  
However, as the chain is compressed to the regime of strong confinement $D\ll R_0$, the scaling relationship for the self-avoiding chain in two dimensions, $R_\parallel^2\sim N^{3/2}$, is observed. 
The variation of scaling exponent $\nu$ from $\nu=1/2$ to $\nu=3/4$ with increasing compression is also found via the scaling relation between $R_{\parallel}$ and $D$ (Eq.\ref{eqn:scaling}).  
Fig.\ref{fig2}C shows the variation of $R_\parallel^2$ versus $D$ for five different values of $N$.

\begin{figure*}[t]
\centering
\includegraphics[width=0.9\textwidth]{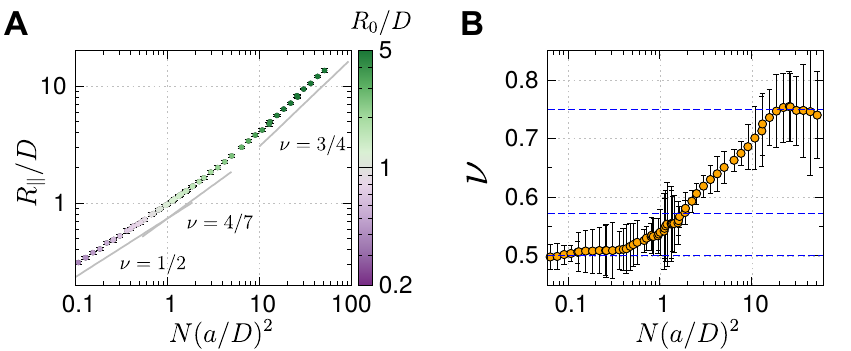}
\caption{(A) All the scattered data depicted in Fig.\ref{fig2} are collapsed onto a single curve using Eq.~\ref{eqn:collapse}. 
The data points are color-coded by the extent of compression ($R_0/D$). 
(B) The scaling exponent $\nu$, calculated by fitting five consecutive data points in the left panel, shows how $\nu$ varies from $\nu=1/2$ to $\nu=3/4$ as a function of $N(a/D)^2$. 
The effect of confinement, which gives rise to the crossover point from $\nu=1/2$ to $\nu> 1/2$, manifests itself at $N(a/D)^2\lesssim 1$. 
Another crossover point to the 2D-SAW ($\nu=3/4$) is found at $N(a/D)^2\gtrsim 10$. 
It is noteworthy that $N(a/D)^2$ corresponds to the number of blobs into which the confined chain is partitioned. }
\label{fig3}
\end{figure*}

In fact, the exponent $\nu$ can be determined in terms of $R_\parallel$, $D$, $a$, and $N$. 
\begin{align}
\nu=\frac{\log{(R_\parallel/D)}}{\log{\left[N(a/D)^{2}\right]}}
\label{eqn:collapse}
\end{align}
Notice that $\Delta F\sim N(a/D)^{2}k_BT$ is the confinement free energy for an ideal chain in slit geometry, and that this free energy, $\Delta F\sim (N/g)k_BT$, corresponds to the energetic cost incurred in partitioning
the original $\Theta$-chain into $N/g$ blobs. 
From Eq.\ref{eqn:blob2},  a relationship between the confinement free energy and the lateral polymer size may be obtained,  
$\beta\Delta F=(R_\parallel/D)^{1/\nu}$. 
Using Eq.\ref{eqn:collapse}, all of the data points from (Fig.\ref{fig2}) collapse onto a single curve (Fig.\ref{fig3}A). 
The slope of $R_\parallel/D$ versus $N(a/D)^2$ defines the scaling exponent $\nu$.  
For a given slit size $D$, the lateral dimension of the confined chain and its scaling exponent change with the extent of compression characterized by the quantity $N(a/D)^2$.  
The size exponent is $\nu=1/2$ for small $N(a/D)^2 < 1$ (for small compression $R_0/D<1$ or before the chain is split into multiple blobs); and plateaus
$\nu=3/4$ for large $N(a/D)^2\gtrsim 10$ (for large compression $R_0/D\gtrsim 5$, or when more than 10 blobs of size $D$ are formed) (Fig.\ref{fig3}A). 
This observation is made more explicit via the scaling exponent $\nu$ calculated by fitting five consecutive data points in Fig.\ref{fig3}A. Fig.\ref{fig3}B shows that there are two regimes where the value of $\nu$ plateaus over the range of compression.


\section{Discussion} 
As the extent of confinement is increased a  $\Theta$-chain  is found to swell.  Its size scales with decreasing $D$ gradually changing from 
$R_0\sim aN^{1/2}$ to $R_\parallel^{\text{SAW}}\sim \tau aN^{3/4}(a/D)^{1/4}$. 
In comparison, 
the evidence for the conformal invariance prediction of $R_\parallel^{\Theta}\sim aN^{4/7}(a/D)^{1/7}$ is weak in Fig.\ref{fig3}B. 
Two crossover points are identified in the scaling exponent $\nu$ as a function of $N(a/D)^2$: 
$N(a/D)^2\sim1$ and $N(a/D)^2\sim 10$ (Fig.\ref{fig3}B). 
The first crossover point $N(a/D)^2\sim1$ arises from the condition when the confinement gap ($D$) is comparable to the size of the chain in free space, $D\sim R_0\simeq aN^{1/2}$. 
Further compression ($D<R_0$) perturbs the chain, which may be visualized as dividing it into multiple blobs. 
The plateau of $\nu=3/4$ is observed for $N(a/D)^2\gtrsim 10$. 
Although there is no clear evidence for the scaling exponent $\nu=4/7$, 
 assuming that the second crossover point is obtained by equating $R_\parallel^\Theta$ and $R_\parallel^{\text{SAW}}$  yields $N(a/D)^2\sim \tau^{-28/5}(a/D)^{7/5}$. 
From this, it could be argued that the second crossover point is sensitive to the solvent quality given in terms of $\tau\left[=(T-T_\Theta)/T_\Theta\right]$. 
Thus, in principle, the second crossover point may be at arbitrarily high confinement as $\tau$ approaches zero.  However, the manifestation of SAW behavior at $N(a/D)^2 \gtrsim  10$ (Fig.\ref{fig3}B) implies either  
that the effect of higher order terms is no longer negligible or that 
the effect of second virial term becomes significant upon confinement. 

In fact, the confinement alone effectively changes the pairwise monomer-monomer interaction which is controlled by the second virial coefficient ($B_2$) \cite{Glandt80JCIS,Yang2015MolPhys,Krekelberg2019JCP}. 
Under slit confinement, $B_2$ changes as a function of $D$.  
For example, for a system comprised of Lennard Jones particles at the Boyle point (for a solution of small particles where the second virial coefficient vanishes), 
$B_2^{\text{LJ}}(D)$ can be derived by approximating the interparticle interaction as a square well potential of range of $\delta$ and depth $\epsilon$ (Fig.\ref{LJ_virial}A) as (see SI for details)
\begin{align}
B^{\text{LJ}}_2(D)=\begin{cases}
\frac{\pi a^3 (D/a)}{2}\left[\left\{\frac{5}{6}+\frac{1}{3}\frac{D}{a}-\frac{1}{6}\left(\frac{D}{a}\right)^2\right\}-(e^{\beta\epsilon}-1)\left\{2\frac{\delta}{a}+\left(\frac{\delta}{a}\right)^2\right\}\right]
& (a<D\leq 2a),\\
\frac{\pi a^3 (D/a) }{2(D/a-1)^2}\left[\left\{\frac{4}{3}\frac{D}{a}\psi(\epsilon,\delta)-\frac{24\psi(\epsilon,\delta)-13}{6}\right\}
-\left(e^{\beta\epsilon}-1\right)\left\{6(\frac{\delta}{a})+5(\frac{\delta}{a})^2+\frac{4}{3}(\frac{\delta}{a})^3\right\}\right]
    & (D\geq 2a).
\end{cases}
\label{eqn:LJsecondvirial}
\end{align}
In the above expression, $\psi(\epsilon,\delta)\left[\equiv 1+(1-e^{\beta\epsilon})\left(\left(1+\frac{\delta}{a}\right)^3-1\right) \right]$ is a factor that takes into account the effect of attraction in the square-well potential (see SI). 
In free space (i.e., $D\rightarrow \infty$) $B_2^{\text{LJ}}(D\rightarrow \infty)=\frac{2\pi}{3}a^3\psi(\epsilon,\delta)$. For the LJ system,  the $\Theta$ condition (i.e., $B_2(\infty)=0$) is gained by setting $\psi(\epsilon,\delta)=0$. 
Remarkably, even for $\psi(\epsilon,\delta)=0$, $B^{\text{LJ}}_2(D)$ still increases from zero to positive values and is maximized to $\sim a^3$ at $D\approx 2a$ as $D$ is decreased (red curve in Fig.\ref{LJ_virial}C). 
In tight confinement ($D\ll R_0$) 
the second virial coefficient, originally tuned to zero, is no longer negligible.  

In summary, in this study the effect of slit confinement on a single $\Theta$-chain in bulk was studied, focusing on the scaling behavior of the chain size with increasing compression. 
Instead of the transition of the scaling exponent from $\nu=1/2$ to $\nu=4/7$, which is expected for a $\Theta$-chain when the dimension is reduced from three to two, our numerical simulation displays the transition from $\nu=1/2$ to $\nu=3/4$ without a clear signature of $\nu=4/7$, which indicates that the solvent quality has effectively been changed from the $\Theta$ to good-solvent condition with increasing compression. 
By highlighting the effect of confining walls on the solvent quality, more specifically on the second virial coefficient, as well as the contribution from the higher order virial terms, this study shows that the tricritical $\Theta$-solvent condition is inherently unstable to reduction of dimensionality. 

Finally, the \emph{variable solvent quality} for polymer under confinement, highlighted in this study, can be related to one of the many factors that determine the stability of a biopolymer under nano-confinement or crowded condition  \cite{zhou2001Biochem,Ziv2005PNAS,Cheung05PNAS,sorin2006JACS,CheungJPCB07,lucent2007PNAS,rasaiah2008ARPC,mittal2008PNAS,inomata2009Nature,lin2012Macromolecules,malik2012JPCB,Ha2015SoftMatter,kim2015SoftMatter,danielsson2015PNAS}.  
Based on a consideration of the polymer conformational entropy, 
it is typically argued that confining a chain molecule alone entropically destabilizes the swollen conformations of chain and facilitates the collapse of the molecule. 
Interestingly, the results from this study based on a single $\Theta$-polymer under confinement suggest that the change in solvent quality from $B_2=0$ to $B_2>0$ engendered by confinement works in the direction that destabilizes the chain. 
Although solvent molecules are implicit, this study offers a possible explanation on why some experiments \cite{malik2012JPCB,inomata2009Nature,danielsson2015PNAS} and simulation studies, which explicitly considered solvent water molecules in conducting folding simulations of a protein molecule inside confinement \cite{lucent2007PNAS,sorin2006JACS}, observe protein destabilization. 

\section{Methods}
$\Theta$-polymer configurations in slit confinement with varying $D$ are generated using molecular simulation of a polymer described below. 
\subsection{Energy potentials} 
The following energy potential was used to generate a $\Theta$ polymer chain. 
\begin{align}
\mathcal{H}_{c}(\{\vec{r}_i\})&=\mathcal{H}_{b}(\{\vec{r}_i\})+\mathcal{H}_{LJ}(\{\vec{r}_i\})\nonumber\\
&=-\frac{k}{2}R_c^2\sum_{i=1}^N{\log{\left(1-\frac{|\vec{r}_i-\vec{r}_{i-1}|^2}{R_c^2}\right)}}\nonumber\\
&+\sum_{i>j}{\epsilon\left[\left\{\left(\frac{a}{r_{ij}}\right)^{12}-2\left(\frac{a}{r_{ij}}\right)^6\right\} -\left\{\left(\frac{a}{r_{c}}\right)^{12}-2\left(\frac{a}{r_{c}}\right)^6\right\} \right]H(r_c-r_{ij})},\end{align}
where $r_{c} = 2.5$ $a$ and $H(\ldots)$ is the Heaviside step function.
In the first term, the finite extensible nonlinear elastic (FENE) potential was used to model the chain connectivity with $k=30$ $k_BT$ and $R_c=1.5$ $a$. 
The second term $\sum_{i>j}u(r_{ij})$ is a shifted Lennard-Jones potential modeling the short-range interaction between monomers. 
The value of $\epsilon(=0.335\text{ }k_BT)$ was chosen such that the second virial coefficient almost vanishes: 
$B_2(T)=\frac{1}{2}\int d{\bf r}(1-e^{-u({\bf r})/k_BT})\simeq \tau a^3$ with $\tau\approx 1.76\times 10^{-3}$. 

Next, the polymer chain generated under $\mathcal{H}_c$ was  confined between a slit of width $D$ by adding a wall potential $\mathcal{H}_w$:  
\begin{align}
\mathcal{H}_{w}(\{z_i\})=\sum_{i=0}^{N}\left[\left(\frac{a}{z_i+\Delta}\right)^{12}-2\left(\frac{a}{z_i+\Delta}\right)^6+1\right]H\left(\Delta-z_i\right)
\end{align}
with $\Delta=a/2$. $\mathcal{H}_w$ models repulsive interactions between the walls and any monomer whose distance from a wall is $z_i\leq a/2$.

\subsection{Simulations}
To efficiently sample the chain configuration in confinement,  the underdamped Langevin equation:   
\begin{align}
m\ddot{\vec{r}}_i=-\zeta\dot{\vec{r}}_i-\nabla_{\vec{r}_i}\mathcal{H}(\{\vec{r_i}\})+\vec{\xi}_i(t)
\end{align}
was integrated, where $\mathcal{H}(\{\vec{r}_i\})=\mathcal{H}_{c}+\mathcal{H}_w$. 
The random force noise, satisfying $\langle\vec{\xi}_i(t)\rangle = 0$ 
and $\langle\vec{\xi}_i(t) \cdot \vec{\xi}_j(t')\rangle = 6\zeta k_{B}T \delta_{ij} \delta(t-t')$, was used to couple the simulated system to Langevin thermostat.
 A small friction coefficient $\zeta=0.1 m/\tau$ and a time step $\delta t=0.005 \tau$ with the characteristic time scale $\tau=(ma^2/\epsilon)^{1/2}$ were chosen.
The whole simulation includes two steps.
(i) A chain which had been pre-equilibrated under the energy potential $\mathcal{H}_{c}$ (i.e., without slit confinement) 
was placed at the center of a large slit with a size of $D=100$ $a$. The slit width, $D$, was then slowly decreased  to $5$ $a$ using a step-size $0.5$ $a$.
At each value of $D$, excessive overlaps between monomers generated from the foregoing stage with larger $D$, 
were eliminated by gradually increasing the short-range repulsive contribution of $\mathcal{H}_c$.
This was achieved by using a modified Lennard-Jones potential term $\mathcal{H}^{\text{mod}}_{LJ}(\{\vec{r}_i\}) = \min\{u_c, \mathcal{H}_{LJ}(\{\vec{r}_i\})\}$ with gradually increasing $u_{c}$.
(ii) The production run was generated for $500$ $\tau$, and chain configurations were collected every $0.5$ $\tau$.
For each combination of $N$ and $D$,  20 replicas were generated starting from different initial configurations and random seeds.
Structural properties averaged over replicas are reported, and their standard deviations are indicated with the error bars. 
Simulations were carried out with the ESPResSo 3.3.1 package \cite{limbach2006espresso}.
\\

{\bf Acknowledgements.} 
We thank the Center for Advanced Computation in KIAS for providing computing resources.
This work was partly supported by the National Research Foundation of Korea (2018R1A2B3001690) (CH). 
\bibliography{mybib1}
\clearpage 

\clearpage

\clearpage

\clearpage

\begin{figure*}[h!]
\centering
\includegraphics[width=0.9\textwidth]{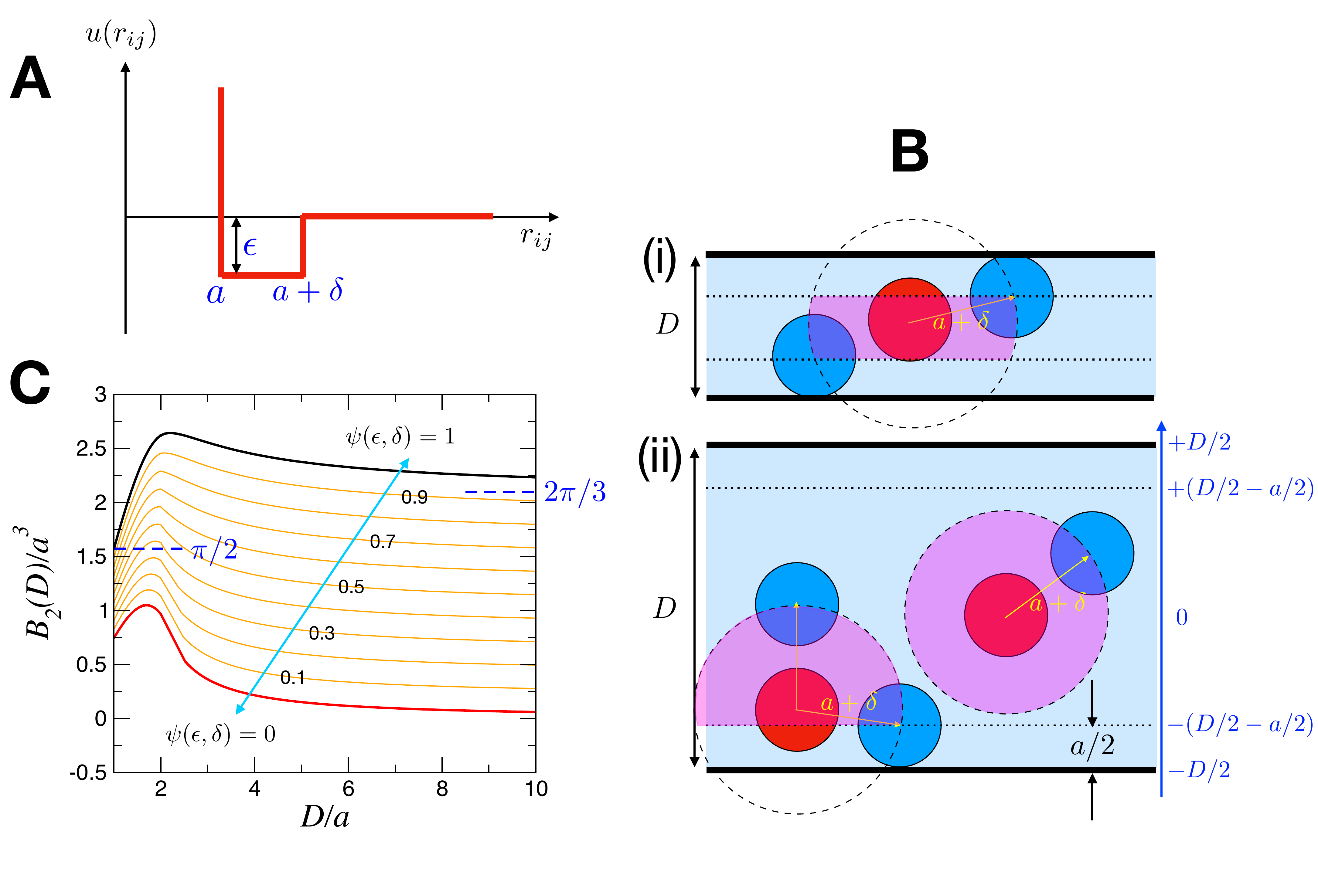}
\caption{The second virial coefficient for Lennard-Jones particles confined in a slit geometry. 
(A) Evaluation of the second virial coefficient for the LJ system is approximated by mapping the system to the one interacting under a square-well potential with interaction range $\delta$ and depth $\epsilon$. 
(B) (i) and (ii) show the configurations for $a\leq D\leq 2a$ and $D\geq 2a$, respectively. 
Illustrated is the effective volume of interaction between red and blue particles interacting with square-well potential depicted in (A). 
The volume of interaction, demarcated using the purple shade of radius $a+\delta$, depends on the position of a reference (red) particle. This cartoon demonstrates the various situations that one has to consider in calculating $B_2(D)$ (see SI).
(C) The $D$-dependent second virial coefficient was calculated with varying $0\leq \psi(\epsilon,\delta)\leq 1$ ($\delta/a=0.5$) by evaluating Eq.\ref{eqn:LJsecondvirial}. 
For $\psi(\epsilon,\delta)=1$, $B_2(D)$ corresponds to that of hard sphere systems confined in a slit (black curve); and $B_2(D)$ for $\psi(\epsilon,\delta)=0$ is the second virial coefficient for the system of LJ particles under $\Theta$ condition (red curve). }
\label{LJ_virial}
\end{figure*}
\clearpage 

\setcounter{equation}{0} \renewcommand{\theequation}{S.\arabic{equation}}

\section{Supporting Information}
\subsection{The second virial coefficient of Lennard Jones particles in slit confinement}
The second virial coefficient for hard-sphere (HS) system amounts to the volume of a spherical particle inaccessible to others. 
That is, the second virial coefficient for homogeneous hard sphere system with diameter $a$, $B_2^{\text{HS}}=\frac{1}{2}\int d{\bf r}(1-e^{-u({\bf{r}})/k_BT})$ with $u({\bf r})=\infty$ for $r<a$ and $0$ for $r\geq a$ is given as $B_2^{\text{HS}}=(2\pi/3)a^3$. 
Geometrical confinement may be regarded as an external field applied to the particles, and it modifies the spatial distribution of a particle at ${\bf r}$ with a factor $g({\bf r})\sim e^{-\beta E({\bf r})}$,  
where $E({\bf r})$ is the potential energy that the particle experiences in the presence of the external field.
The second virial coefficient of hard sphere system in slit confinement is given as \cite{Krekelberg2019JCP}
\begin{align}
B_2&=-\frac{V}{2}\frac{\int d{\bf r}_1g({\bf r}_1)\int d{\bf r}_2g({\bf r}_2)f({\bf r}_1,{\bf r}_2)}{\int d{\bf r}_1g({\bf r}_1)\int d{\bf r}_2g({\bf r}_2)}. 
\end{align}
In this expression, the Mayer function is $f({\bf r}_1,{\bf r}_2)=e^{-\beta u(r_{12})}-1=H(|r_1-r_2|-a)-1$ for hard spheres where $H(\ldots)$ is the Heaviside step function. 
The integral $g({\bf r}_1)\int d{\bf r}_2g({\bf r}_2)f({\bf r}_1,{\bf r}_2)$ 
corresponds to the volume around a particle at ${\bf r}_1$ that is inaccessible to other particles under the constraint that all the particles are confined between the slit. 
The integral $\int d{\bf r}g({\bf r})=A(D-a)$ denotes the volume accessible for a sphere of diameter $a$ inside the slit of width $D$ and area $A$. 
Since the inaccessible volume around the particle varies with its location along the $z$ axis, $B_2$ can be written in the following form.  
\begin{align}
B_2(D)=\frac{AD}{2}\frac{A\int_{-D/2+a/2}^{D/2-a/2} dzV(z)}{A^2(D-a)^2}
\label{eqn:B2}
\end{align}

Next, in order to evaluate the second virial coefficient,  a system of Lennard-Jones particles may be approximated by a square well potential of depth $\epsilon$ and width $\delta$ (see Fig.\ref{LJ_virial}A). 
In free space, the inaccessible volume calculated above for hard sphere system is adjusted by a factor $\psi(\epsilon,\delta)
$ as 
\begin{align}
\int d{\bf r}(1-e^{-\beta u({\bf r})})&=4\pi\int_0^a r^2dr+4\pi\int_a^{a+\delta} r^2dr(1-e^{\beta \epsilon})\nonumber\\
&=\frac{4\pi}{3}a^3\underbrace{\left[1+(1-e^{\beta\epsilon})\left(\left(1+\frac{\delta}{a}\right)^3-1\right)\right]}_{\equiv \psi(\epsilon,\delta)}
\end{align}
The effective volume element $V(z)$ for LJ in Eq.\ref{eqn:B2} is calculated as (see Fig.\ref{LJ_virial}B):

(i) $a\leq D\leq 2a$, 
\begin{align}
V_{\text{LJ}}^<(z)
&=\pi\int^{\frac{D}{2}-\frac{a}{2}-z}_{-\frac{D}{2}+\frac{a}{2}-z}dy(a^2-y^2)
+\pi(1-e^{\beta \epsilon})\int_{-\frac{D}{2}+\frac{a}{2}-z}^{\frac{D}{2}-\frac{a}{2}-z}dy\left[(a+\delta)^2-y^2\right]\nonumber\\
&-\pi(1-e^{\beta\epsilon})\int_{-\frac{D}{2}+\frac{a}{2}-z}^{\frac{D}{2}-\frac{a}{2}-z}dy(a^2-y^2)
\nonumber
\end{align}

(ii) $D\geq 2a$, 
\begin{itemize}
\item For $\frac{D}{2}-\frac{3a}{2}\leq z\leq \frac{D}{2}-\frac{a}{2}$
\begin{align}
V^{>,1}_{\text{LJ}}(z)&=\pi e^{\beta\epsilon}\int^{D/2-a/2-z}_{-a}dy(a^2-y^2)+\pi (1-e^{\beta\epsilon})\int^{D/2-a/2-z}_{-(a+\delta)}dy((a+\delta)^2-y^2)\nonumber
\end{align}
\item For $-\frac{D}{2}+\frac{3a}{2}\leq z\leq \frac{D}{2}-\frac{3a}{2}$
\begin{align}
V^{>,2}_{\text{LJ}}(z)&=\pi e^{\beta\epsilon}\int^{a}_{-a}dy(a^2-y^2)
+\pi(1-e^{\beta\epsilon}) \int^{a+\delta}_{-a-\delta}dy((a+\delta)^2-y^2)\nonumber\\
&=\frac{4\pi}{3}a^3\psi(\epsilon,\delta)\nonumber
\end{align}
\item For $-\frac{D}{2}+\frac{a}{2}\leq z\leq -\frac{D}{2}+\frac{3a}{2}$
\begin{align}
V^{>,3}_{\text{LJ}}(z)&=\pi e^{\beta\epsilon}\int^{a}_{-D/2+a/2-z}dy(a^2-y^2)
+\pi(1-e^{\beta\epsilon}) \int^{a+\delta}_{-D/2+a/2-z}dy((a+\delta)^2-y^2)\nonumber
\end{align}
\end{itemize}
Then, $B_2(D)$ for LJ systems is obtained by calculating 
$B_2^{\text{LJ}}(D)=\frac{D}{2(D-a)^2}\int^{D/2-a/2}_{-D/2+a/2}V_{\text{LJ}}(z)dz$. 

 Eq.\ref{eqn:B2} is evaluated with the explicit forms of $V_{\text{LJ}}^<(z)$ and $V_{\text{LJ}}^{>,i}(z)$ given above: 

(i) For $a\leq D\leq 2a$
\begin{align}
B_2^{\text{LJ}}(D)=\frac{D}{2(D-a)^2}\int_{-D/2+a/2}^{D/2-a/2}dzV_{\text{LJ}}^{<}(z)
\end{align}

(ii) For $D\geq 2a$
\begin{align}
B_2^{\text{LJ}}(D)=\frac{D}{2(D-a)^2}\left(\int_{-D/2+a/2}^{-D/2+3a/2}dzV^{>,1}_{\text{LJ}}(z)+\int_{-D/2+3a/2}^{D/2-3a/2}V^{>,2}_{\text{LJ}}(z)dz+\int_{D/2-3a/2}^{D/2-a/2}V^{>,3}_{\text{LJ}}(z)dz\right)
\end{align}
These enable obtaining the expression given in Eq.\ref{eqn:LJsecondvirial}.

The condition of either $\psi(\epsilon,\delta)\rightarrow 1$ ($\epsilon=0$ or $e^{\beta\epsilon}=1$) or $\delta/a=0$ reduces the second virial coefficient of LJ system into that of hard-sphere systems \cite{Yang2015MolPhys}: 
\begin{align}
B^{\text{LJ}}_2(D)\rightarrow B^{\text{HS}}_2(D)=\begin{cases}
\frac{\pi a^3(D/a)}{2}\left[\frac{5}{6}+\frac{1}{3}\frac{D}{a}-\frac{1}{6}\left(\frac{D}{a}\right)^2\right] & (a<D\leq 2a),\\
\frac{\pi a^3(D/a)}{2(D/a-1)^2}\left[\frac{4}{3}\frac{D}{a}-\frac{11}{6}\right] & (D\geq 2a).
\end{cases}
\label{eqn:secondvirial}
\end{align}
$B_2^{\text{HS}}(D)$ changes its value from $B^{\text{HS}}_2(D\rightarrow\infty)=(2\pi/3)a^3$ in free space to $B^{\text{HS}}_2(D\rightarrow a)=(\pi/2)a^3$ in tightly confined space. 
It is interesting to note that both $B^{\text{HS}}_2(D)$ and $B^{\text{LJ}}_2(D)$ display non-monotonic variation with $D$, maximized at $D\approx 2a$ (see black curve in Fig.\ref{LJ_virial}C).   

\end{document}